# Superhump periods in the UGSU-type dwarf nova SDSSp J082409.73+493124.4

David Boyd, Jeremy Shears, Robert Koff

**Abstract**

We present observations and analysis of the first reported superoutburst of the dwarf nova SDSSp J082409.73+493124.4 during February/March 2007. From a maximum observed magnitude of 15.4C it declined at 0.09 mag $d^{-1}$ for 7 days, flattened out around magnitude 16 for a further 5 days and then returned rapidly to quiescence at magnitude 19.4. The flattening of the light curve late in the outburst was not associated with a re-growth of superhumps. For the first 5 days we observed common superhumps with period 0.06954(5) d, thus confirming its classification as a UGSU-type dwarf nova. This was followed by a phase transition to late superhumps with period 0.06921(6) d. We found a small but persistent signal at 0.0687(6) d which we interpret as the orbital period.

**Introduction**

Cataclysmic variables (CVs) are close binary systems in which a white dwarf is gathering material via an accretion disc from its companion, typically a late main-sequence star [1]. Dwarf novae are a sub-class of CVs which undergo quasi-periodic outbursts from their quiescent state caused by a thermal instability in the disc during which they may brighten by several magnitudes for a few days. In some cases these outbursts may be brighter and more prolonged and may exhibit periodic modulations called superhumps arising from eccentricity in the accretion disc. These longer outbursts are referred to as superoutbursts and indicate that the dwarf nova belongs to the sub-type UGSU, named after the two prototype variables which define this type and sub-type: U Geminorum and SU Ursa Majoris.

SDSSp J082409.73+493124.4 (hereafter referred to as SDSS0824), which is in the constellation Lynx, was first recognised as a dwarf nova by the Sloan Digital Sky Survey (SDSS) in November 2000 at which time its g* magnitude (similar to visual) was 19.28 and its r* magnitude was 18.84 [2]. This survey identifies CVs from their strong hydrogen Balmer and helium emission lines which typically signify the accretion process. The same paper also reported observations of the object on two days by the Lowell Observatory Near-Earth Object Search (LONEOS) project which recorded SDSS0824 in the magnitude range 17.60 – 17.75 with error 0.1, calibrated from USNO-A2.0 red magnitudes. Vizier [3] gives a USNO-A2.0 catalogue red magnitude of 18.8. There is therefore some evidence of variability. Downes et al. [4] lists the object at magnitude 19.3 and indicates that no period is known.

The AAVSO database [5] reveals only one positive observation of this object prior to the outburst reported here: an unfiltered CCD measurement on 2006 November 4 at magnitude 15.9. Unfortunately there were no corroborating observations at that time. With the possible exception of this single observation, there have been no previous observations of SDSS0824 in outburst. Prior to 2006 November the object received virtually no coverage so we do not know its supercycle length.

**Superoutburst in February/March 2007**

JS reported SDSS0824 in outburst on 2007 Feb 28.944 at an unfiltered (Clear) magnitude of 15.7C [6] (subsequently recalibrated to 15.4C). He observed modulation in the light curve which was confirmed by DB the following night when a clear superhump signal was present. This confirmed the UGSU classification of SDSS0824. Over the next 20 days, we carried out 20 CCD photometry runs comprising 61.6 hours of integration and 2605 images including 12 time-series longer than 1.8 hours. The proximity of the moon during the early stages of the outburst (ameliorated briefly during the total lunar eclipse on March 3), plus poor observing conditions in the UK for much of the outburst, resulted in high background levels in many of the images. All data was taken unfiltered to maximise signal-to-noise. Figure 1 shows SDSS0824 in outburst on March 3 (during the lunar eclipse).

All images were dark-subtracted and flat-fielded before being measured using differential aperture photometry relative to nearby comparison stars. We used comparison star magnitudes measured by Henden [7] and verified internal consistency between the magnitudes of the comparison stars we used.

A log of observations is given in Table 1 and the equipment used by each observer is listed in Table 2. Julian Dates given in the paper are in the truncated form JD = JD – 2454000. The combined light curve for the superoutburst is shown in Figure 2 where each run is represented by its mean magnitude and time. The initial observation and some of the longer runs are shown in Figure 3, all plotted at the same magnitude and time scale. From JD 160 to 167 the outburst declined at 0.09 mag $d^{-1}$. It then flattened out around magnitude 16.0 for a further 5 days before dropping rapidly back to quiescence at magnitude 19.4. The amplitude of the outburst was approximately 4 magnitudes. We examined the amplitude of the superhump signal in the individual light curves but saw no evidence of a re-growth of superhumps during the period when the light curve flattened towards the end of the outburst.

**Superhump maximum analysis**

Times of superhump maximum were obtained by fitting a quadratic function to 22 sufficiently well-defined superhump maxima. Following a preliminary assignment of superhump cycle numbers to these maxima, an analysis of the times of maximum indicated that the superhump period appeared to remain constant for the first 5 days, sometime after which it changed. We obtained the following linear superhump maximum ephemeris for the interval JD 160 to 164:

HJD 2454160.461(2) + 0.06954(5) * E

The observed minus calculated (O–C) residuals relative to this ephemeris are plotted in Figure 4 and listed in Table 3 along with the measured times of superhump maximum and their assigned superhump cycle numbers. Following accepted practice, we interpret the interval JD 160 to 164 as being a common superhump regime. Sometime between JD 165 and 167 the superhump phase changed by at least 0.25 and possibly by more than that. This is consistent with a transition to late superhumps from JD 167 onwards. The negative slope of the linear fit to the late superhump times of maximum in Figure 4 indicates that the late superhump period was shorter than the common superhump period. We obtained the late superhump maximum ephemeris:

HJD 2454160.513(8) + 0.06921(6) * E

Assuming constant periods for the common and late superhumps, we find a common superhump period of 0.06954(5) d and a late superhump period of 0.06921(6) d. For a discussion of common and late superhumps, see for example [1].

**Period analysis**

We selected light curves which were longer than 1.8 hr to ensure they covered at least one complete superhump cycle and subtracted linear trends from these individually. We carried out separate period analyses on the data in the common superhump regime (JD 160 – 164) and in the late superhump regime (JD 167 onwards) using the Lomb-Scargle method in Peranso [8].

For the common superhump regime, we obtained the power spectrum shown in Figure 5. Interpreting the main peak as the superhump period gives a common superhump period of 0.0697(4) d. For the late superhump regime, the power spectrum is shown in Figure 6 and gives a late superhump period of 0.0695(4) d. These periods are consistent with those found from analysing times of superhump maxima. The strong +/-0.5 d aliases in the common superhump spectrum are expected because of the 2-day separation of several of the runs. Phase diagrams for common and late superhumps are shown in Figures 7 and 8 respectively. The mean peak-to-peak amplitude of common superhumps is ~0.13 mag and for late superhumps is ~0.08 mag.

**Orbital period and superhump period excess**

We removed the superhump signal and examined the residual power spectrum in both superhump regimes for evidence of an orbital signal. The strongest residual signal in the common superhump regime was at 0.0687(7) d and in the late superhump regime was at 0.0687(4) d. Although these were both weak signals, their common value gives us confidence that they probably represent the orbital signal. Assuming this to be the case gives an orbital period of 0.0687(6) d = 98.9(8) min.

Dillon et al. [9] have obtained two estimates of the orbital period from quiescent photometry: 98.133(56) min and 94.987(54) min, with the shorter time slightly preferred. Their photometry exhibited the double-humped morphology typically observed in those short-period low mass transfer systems.

Our result favours their longer period. This would give a superhump period excess, $\varepsilon$, of 2.0%, while their shorter period would give 5.4%. Considering the distribution of published values of $\varepsilon$, for example in Figure 6.6 in [1] or Figure 6 in [10], the lower value is more consistent with values of $\varepsilon$ for other dwarf novae with similar orbital periods. The higher value would be well outside the range of current published values. This confirms our conclusion that the correct orbital period is 98.133(56) min.

**Comparison with other UGSU dwarf novae**

Flattening or re-brightening of the light curve late in a superoutburst as seen in SDSS0824 has been observed in other UGSU dwarf novae, for example HO Del [11] and HV Vir [12]. This behaviour has also been associated with a re-growth in superhump amplitude in some systems such as V1028 Cyg [13]. Table 3 in [11] indicates that, with one exception (RZ Sge), re-growth of superhumps in UGSU systems is always associated with late re-brightening of the light curve, although the reverse is not true. Figures 11 and 12 in [11] show that there is a tendency for systems with late re-brightening to have superhump periods shorter than 0.07 d and supercycles longer than about 250 d. The measured superhump period and single observed superoutburst of SDSS0824 are consistent with this.

**Astrometry of SDSSp J082409.73+493124.4**

We obtained the following mean position from 6 good quality images of the dwarf nova in outburst using Astrometrica [14] and the USNO CCD Astrograph Catalog release 2:

RA 8h 24m 9.71s +/- 0.15s, Dec 49º 31' 24.85" +/- 0.11" (J2000)

We note there is a 0.4" discrepancy in declination compared with the position RA 8h 24m 9.72s, Dec 49º 31' 24.4" (J2000) listed in the SDSS DR5 database [15].

**Conclusion**

We have observed and analysed the first reported superoutburst of the dwarf nova SDSSp J082409.73+493124.4 during February/March 2007. From a maximum observed magnitude of 15.4C it declined at 0.09 mag $d^{-1}$ for 7 days, flattened out around magnitude 16 for a further 5 days and then returned rapidly to quiescence at magnitude 19.4. We therefore observed the whole outburst. The flattening of the light curve observed late in the outburst was not associated with a re-growth of the superhumps as has been seen in some UGSU systems. During the first 5 days of the outburst, common superhumps were seen with a period of 0.06954(5) d confirming its classification as a UGSU-type dwarf nova. There was then a superhump phase transition to a late superhump regime with a superhump period of 0.06921(6) d. After removing both common and late superhump signals we found a small but persistent orbital signal at 0.0687(6) d = 98.9(8) min. This favours the longer orbital period of 98.133(56) min found by Dhillon et al. [9].

**Acknowledgements**

We acknowledge with thanks access to variable star observations from the AAVSO International Database contributed by observers worldwide. We also thank Dr Boris Gaensicke for information about measurements of the orbital period and both referees for their helpful comments.

**Addresses**

DB: 5 Silver Lane, West Challow, Wantage, Oxon, OX12 9TX, UK
   [drsboyd@dsl.pipex.com]
JS: "Pemberton", School Lane, Bunbury, Tarporley, Cheshire, CW6 9NR, UK
   [bunburyobservatory@hotmail.com]
RK: 980 Antelope Drive West, Bennett, CO 80102, USA
   [bob@AntelopeHillsObservatory.org]

**References**

1. Hellier C., Cataclysmic Variable Stars: How and why they vary, Springer-Verlag, 2001
2. Szkody P. et al., Astron. J., **123**, 430 (2002)
3. Vizier Catalogue Service, http://vizier.u-strasbg.fr/
4. Downes R. A. et al., http://archive.stsci.edu/prepds/cvcat/index.html
5. AAVSO, http://aavso.org
6. Shears J., BAAVSS Alert Group message 1051, http://tech.groups.yahoo.com/group/baavss-alert/message/1051
7. Henden A., ftp://ftp.aavso.org/public/calib/p443f360.dat
8. Vanmunster T., Peranso, http://www.peranso.com
9. Dillon M. et al., Mon. Not. R. Astron. Soc., to be submitted (2007)
10. Pearson K.J., Mon. Not. R. Astron. Soc., **371**, 235 (2006)
11. Kato T., Nogami D., Moilanen M. and Yamaoka H., Publ. Astron. Soc. Japan, **55**, 989 (2003)
12. Ishioka R. et al., Publ. Astron. Soc. Japan, **55**, 683 (2003)
13. Baba H. et al., Publ. Astron. Soc. Japan, **52**, 429 (2000)
14. Raab H., Astrometrica, http://www.astrometrica.at
15. Sloan Digital Sky Server, http://cas.sdss.org/dr5/en/

| Date in 2007 (UT) | Start time (JD) 2454000+ | Duration (h) | No. of images | Mean mag (C) | Std dev mag (C) | Observer |
|---|---|---|---|---|---|---|
| Feb 28 | 160.44552 | 0.8 | 45 | 15.44 | 0.10 | JS |
| Mar 1 | 161.28702 | 7.0 | 368 | 15.56 | 0.06 | DB |
| Mar 1 | 161.29583 | 6.6 | 272 | 15.64 | 0.13 | JS |
| Mar 3 | 163.38936 | 4.6 | 245 | 15.83 | 0.14 | JS |
| Mar 3 | 163.39837 | 2.9 | 170 | 15.73 | 0.04 | DB |
| Mar 4 | 163.57017 | 8.0 | 207 | 15.65 | 0.06 | RK |
| Mar 5 | 164.56985 | 5.7 | 146 | 15.78 | 0.04 | RK |
| Mar 6 | 166.49699 | 0.6 | 26 | 16.06 | 0.06 | DB |
| Mar 7 | 167.31907 | 0.6 | 34 | 16.15 | 0.16 | JS |
| Mar 7 | 167.39456 | 1.9 | 70 | 16.10 | 0.08 | DB |
| Mar 8 | 167.73148 | 3.0 | 76 | 15.95 | 0.04 | RK |
| Mar 9 | 168.73415 | 3.8 | 100 | 15.91 | 0.04 | RK |
| Mar 9 | 169.30314 | 5.5 | 282 | 16.09 | 0.04 | DB |
| Mar 9 | 169.30535 | 0.0 | 2 | 16.02 | 0.07 | JS |
| Mar 10 | 170.36563 | 4.7 | 260 | 16.19 | 0.03 | DB |
| Mar 12 | 171.79379 | 1.4 | 39 | 16.10 | 0.04 | RK |
| Mar 12 | 172.34764 | 1.0 | 47 | 16.41 | 0.07 | DB |
| Mar 13 | 173.44162 | 3.0 | 188 | 17.01 | 0.06 | DB |
| Mar 14 | 174.36586 | 0.2 | 14 | 17.87 | 0.15 | DB |
| Mar 20 | 180.41731 | 0.2 | 14 | 19.38 | 0.31 | DB |

Table 1: Log of observations.

| Observer | Aperture (m) | Telescope type | CCD camera |
|---|---|---|---|
| DB | 0.36 | SCT | Starlight Xpress SXV-H9 |
| JS | 0.10 | Fluorite refractor | Starlight Xpress SXV-M7 |
| RK | 0.25 | SCT | Apogee AP-47 |

Table 2: Equipment used.

| Date in 2007 (UT) | Time of maximum (HJD) 2454000+ | Superhump cycle no | O–C (cycles) |
| --- | --- | --- | --- |
| Feb 28 | 160.46358 | 0 | 0.035 |
| Mar 01 | 161.36430 | 13 | -0.012 |
| Mar 01 | 161.43506 | 14 | 0.005 |
| Mar 02 | 161.57139 | 16 | -0.034 |
| Mar 03 | 163.45018 | 43 | -0.018 |
| Mar 04 | 163.51899 | 44 | -0.028 |
| Mar 04 | 163.58942 | 45 | -0.015 |
| Mar 04 | 163.65881 | 46 | -0.018 |
| Mar 04 | 163.73223 | 47 | 0.038 |
| Mar 04 | 163.80305 | 48 | 0.057 |
| Mar 04 | 163.86739 | 49 | -0.018 |
| Mar 05 | 164.63419 | 60 | 0.008 |
| Mar 08 | 167.77860 | 105 | 0.225 |
| Mar 09 | 168.74764 | 119 | 0.159 |
| Mar 09 | 168.81854 | 120 | 0.179 |
| Mar 09 | 168.88736 | 121 | 0.169 |
| Mar 09 | 169.36972 | 128 | 0.105 |
| Mar 09 | 169.44292 | 129 | 0.158 |
| Mar 10 | 169.50932 | 130 | 0.112 |
| Mar 10 | 170.40944 | 143 | 0.056 |
| Mar 10 | 170.47796 | 144 | 0.041 |
| Mar 11 | 170.54702 | 145 | 0.034 |

Table 3: Times of superhump maximum.

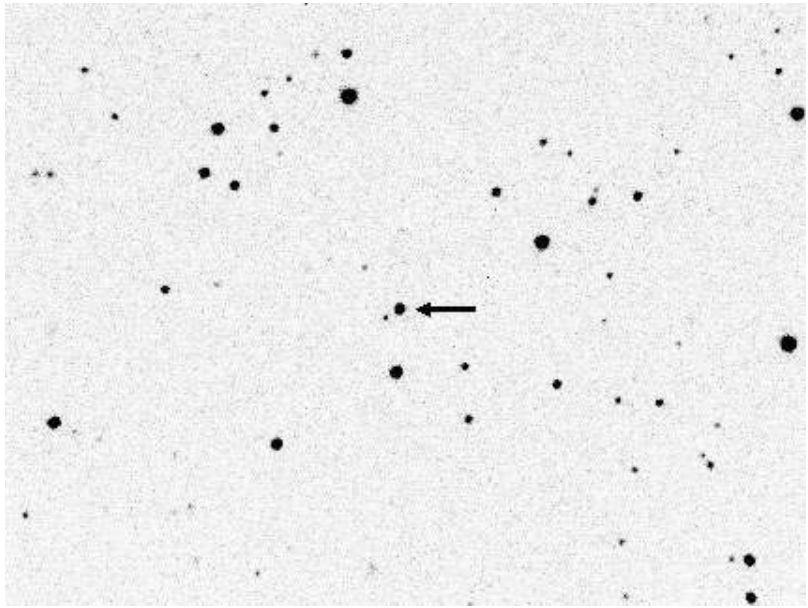

Figure 1: SDSS0824 in outburst on March 3, field 10' wide, north at top.

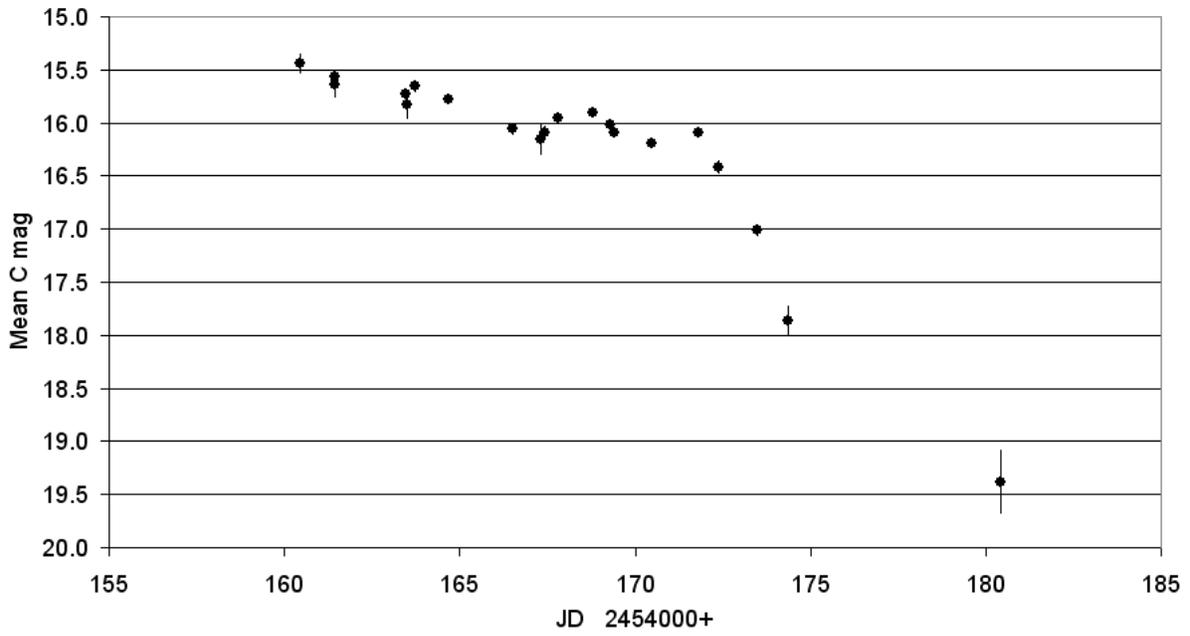

Figure 2: SDSS0824 superoutburst light curve.

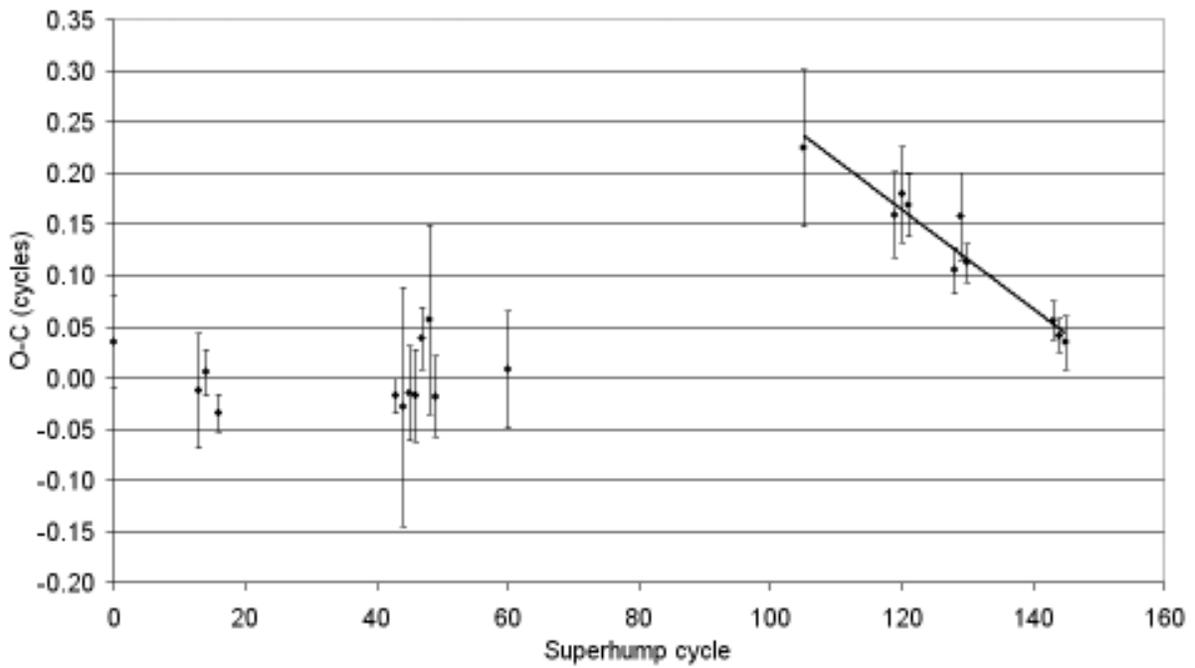

Figure 4: O–C residuals to the common superhump ephemeris.

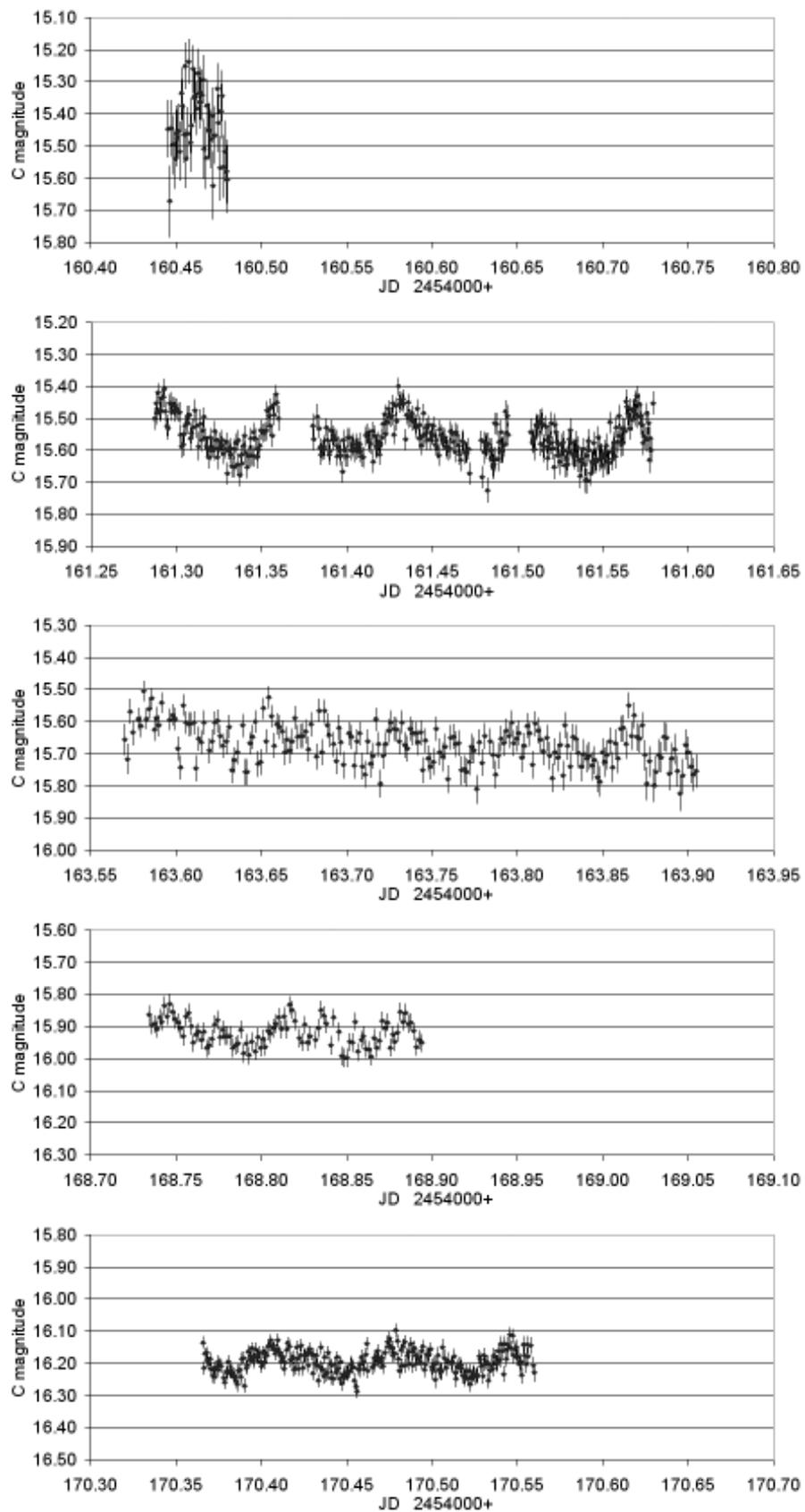

Figure 3: Initial observation and some of the longer runs, all at the same magnitude and time scale.

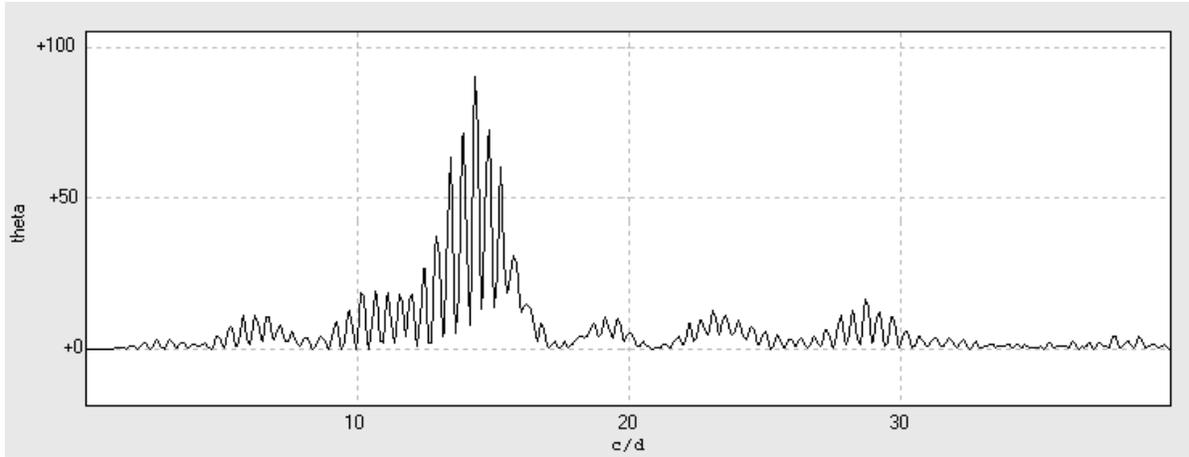

Figure 5: Power spectrum for the common superhump regime (JD 160 – 164).

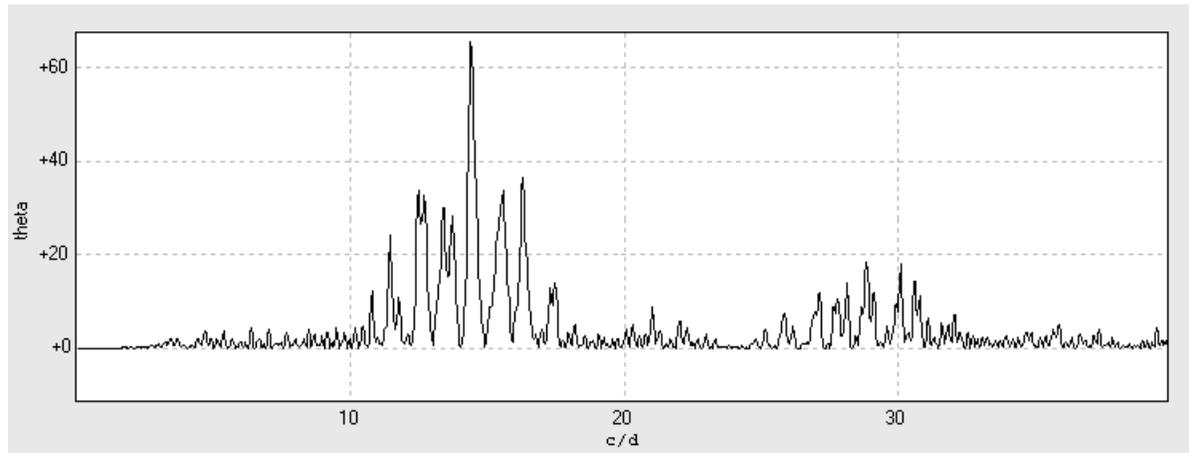

Figure 6: Power spectrum for the late superhump regime (JD 167 onwards).

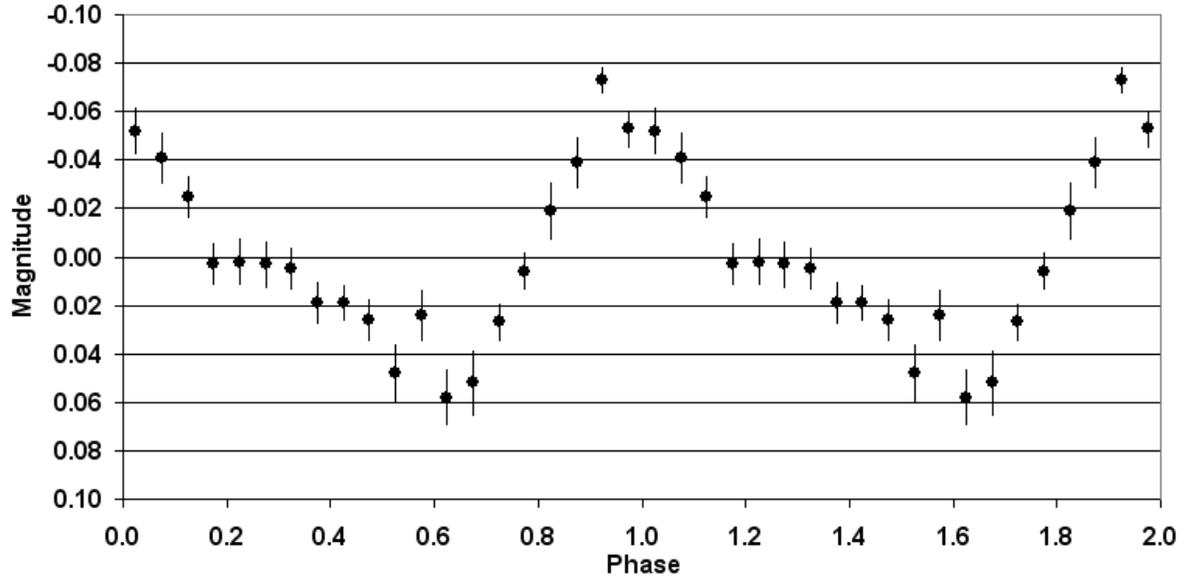

Figure 7: Phase diagram for common superhumps (2 cycles).

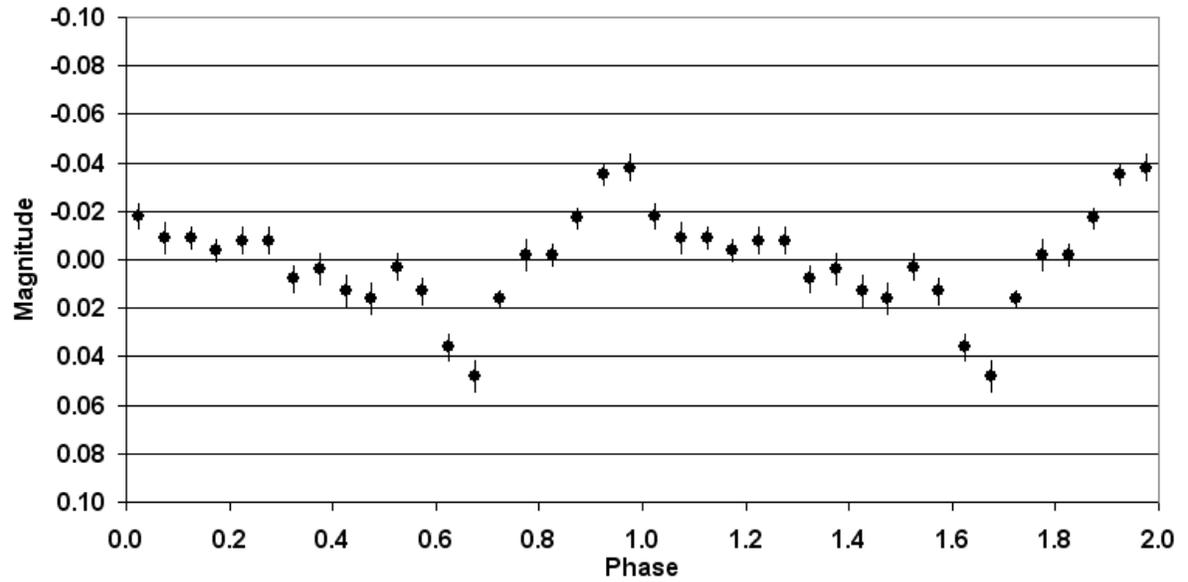

Figure 8: Phase diagram for late superhumps (2 cycles).